# Taking Heisenberg's Potentia Seriously

R. E. Kastner[1]
Stuart Kauffman[2]
Michael Epperson[3]

March 16, 2018

It is argued that quantum theory is best understood as requiring an ontological dualism of *res extensa* and *res potentia*, where the latter is understood per Heisenberg's original proposal, and the former is roughly equivalent to Descartes' 'extended substance.' However, this is not a dualism of mutually exclusive substances in the classical Cartesian sense, and therefore does not inherit the infamous 'mind-body' problem. Rather, res potentia and res extensa are understood as mutually implicative ontological extants that serve to explain the key conceptual challenges of quantum theory; in particular, nonlocality, entanglement, null measurements, and wave function collapse. It is shown that a natural account of these quantum perplexities emerges, along with a need to reassess our usual ontological commitments involving the nature of space and time.

## I. Introduction and Background

It is now well-established, via the violation of the various Bell Inequalities, that Nature at the quantum level entails a form of nonlocality such that it is not possible to account for phenomena in terms of local common causes. Shimony (2017) provides a comprehensive review of this topic. Many researchers have explored, and continue to explore, various ways of retaining some form of pseudo-classical locality in the face of these features of quantum theory. Among these are:

- Time-Symmetric Hidden Variables theories (e.g., Price and Wharton, 2015 , Sutherland, 2017)

[1] Foundations of Physics Group, University of Maryland, College Park
[2] Institute for Systems Biology, Seattle, and Professor Emeritus, Dept. of Biochemistry and Biophysics, University of Pennsylvania
[3] Center for Philosophy and the Natural Sciences, College of Natural Sciences and Mathematics, California State University Sacramento

- Many Worlds (Everettian) Interpretations

- "Quantum Bayesianism" or "Qbism" (Fuchs, Mermin, Schack, 2014)

- The "Bohmian" theory (first proposed in Bohm, 1952)

Of course, the Bohmian theory does have nonlocal influences, but seeks to retain at least pseudo-classical localizability via its postulated corpuscles. Meanwhile, instrumentalism simply evades the ontological challenges posed by evidently nonlocal quantum behavior. Qbism (described by its founders as a 'normative' rather than descriptive interpretation) claims to retain locality, but does so through a maneuver in which even a radically nonlocal theory (not quantum theory, but a hypothetical one that allows explicit faster-than-light signaling) must also be deemed 'local,' as shown by Henson (2015). We believe that this constitutes a *reductio ad absurdum* of the Qbism 'locality' argument, showing that it fails to establish locality in any physically meaningful sense. Of course, we argue here that locality in light of quantum theory is not something that we should be trying to retain, anyway: the quantum 'spookiness,' seen as something needing to be suppressed or eliminated in the above approaches, may actually be an important clue to a richer ontology of the world than has been previously suspected.[1] This is what we explore herein.

Thus, we suggest here that, rather than retreating into instrumentalism or trying to 'save locality' (either partially or fully) by adding various *ad hoc* quantities to the formalism (i.e., hidden variables), it is worthwhile to consider the possibility that the world is indeed nonlocal at the quantum level, and to seek a fruitful ontological explanation. In this regard, we want to focus our attention on Heisenberg's original suggestion that quantum entities can be understood as a form of Aristotle's 'potentia.' For Heisenberg, potentiae are not merely epistemic, statistical approximations of an underlying veiled reality of predetermined facts; rather, potentiae are ontologically fundamental constituents of nature. They are things "standing in the middle between the idea of an event and the actual event, a strange kind of physical reality just in the middle between possibility and reality" (Heisenberg 1958, 41). Elsewhere, Heisenberg suggests that one consider quantum mechanical probabilities as "a new kind of 'objective' physical reality. This probability concept is closely related to the concept of natural philosophy of the ancients such as Aristotle; it is, to a certain extent, a transformation of the old 'potentia' concept from a qualitative to a quantitative idea" (Heisenberg 1955, 12). It is worth nothing that Shimony (e.g.,

---

[1] It seems important to note here that historically, anomalies that seemed absurd, unacceptable, or unexplainable given a particular metaphysical model have always turned out to be natural and understandable with the adoption of a new model—and that constituted scientific progress. A case in point is the anomalous motions of the planets in an Earth-centered ontology. Without meaning any disrespect to the esteemed authors cited above, it does seem to us that trying to retain our usual ontological commitments by 'tweaking' the basic theory with the addition of *ad hoc* quantities like hidden variables is somewhat akin to constructing Ptolemaic epicycles. In fact, as his colleague Basil Hiley has pointed out (private communications), Bohm himself abandoned the 1952 model and pursued other realist approaches to understanding quantum theory, a quest with which the current authors are sympathetic. We discuss this methodological issue, of concern not only to the present

1997) also considered this concept in connection with solving interpretational challenges of quantum theory (although he largely supposed that the mental was a necessary aspect in the conversion of potentiality to actuality, with which the present authors differ).

As a further prelude to the metaphysical picture being considered herein, first recall Descartes' dualism of *res cogitans* (purely mental substance) and *res extensa* (purely physical substance) as mutually exclusive counterparts. While seemingly plausible, it encountered notorious difficulties in view of the fact that mutually exclusive substances, by definition, cannot be integrated, thus leading to the 'mind/body' problem. Of course, there were many and varied responses to this problem, which we will not enter into here; but a common response among physical scientists has been simply to reject *res cogitans* and to assume that only *res extensa* exists. This would seem to be a serviceable approach for classical physics (even if subject to criticism in view of the 'hard problem of consciousness' (Chalmers 1995)). However, with the advent of quantum physics, new explanatory challenges arise that may be fruitfully met by considering a richer ontology.

We thus propose a new kind of ontological duality as an alternative to the dualism of Descartes: in addition to *res extensa*, we suggest, with Heisenberg, what may be called *res potentia*. We will argue that admitting the concept of potentia into our ontology is fruitful, in that it can provide an account of the otherwise mysterious nonlocal phenomena of quantum physics and at least three other related mysteries ('wave function collapse'; loss of interference on which-way information; 'null measurement'), without requiring any change to the theory itself. This new duality omits Descartes' *res cogitans*. In addition, it should be noted that with respect to quantum mechanics, *res potentia* is not itself a separate or separable substance that can be ontologically abstracted from *res extensa* (i.e., neither can be coherently defined without reference to the other, in contrast to *res extensa* and *res cogitans* in the Cartesian scheme). Thus, in the framework proposed herein, actuality and potentiality will not be related as a *dualism of mutually exclusive concepts*, but rather a *duality of mutually implicative concepts*.[2]

As indicated by the term '*res*,' we do conceive of *res potentia* as an ontological extant in the same sense that *res extensa* is typically conceived—i.e. as 'substance,' but in the more general, Aristotelian sense, where substance does not necessarily entail conflation with the concept of physical matter, but is rather merely "the essence of a thing . . . what it is said to be *in respect of itself*" (*Metaphysics* Z.4. 1029b14). Substance, in this regard, is the essence and definition of a thing, such that the things defined (here, actuality and potentiality) are not further reducible, physically or conceptually. Thus, in the framework proposed herein, *res extensa* and *res potentia* are the two fundamental, mutually implicative ontological constituents of nature at the quantum mechanical level. More specifically, they are mutually implicative constituents of every quantum measurement event.[3] Therefore, our thesis does not inherit the mind-body problem of Cartesian dualism, in which two fundamentally different, *mutually exclusive*, substances have no way of interacting. Two of us, Kauffman and Epperson, have addressed the relevance for the mind-body problem elsewhere (Kauffman 2016, Chapter 8; Epperson 2009, 344-353).

---

[2] For further elaboration, see Epperson (2013, 4-10) and Eastman (2003, 14-30).

Thus, the new metaphysical picture, which we will argue is supported by quantum theory and its empirical success, consists of an ontological duality: *res potentia* and *res extensa*. In quantum mechanics, these are exemplified, respectively, as systems in pure states (i.e., rays in Hilbert space) and *actual* system outcomes, which are not represented by pure states, but instead by projection operators corresponding to the actual outcome. In this way, the evaluation of an observable via a quantum measurement event entails the actualization of one of the potential outcomes inherent in a pure state (i.e. a given pure state embodies many potential outcomes). It is a fundamental feature of quantum mechanics that the object of observation is always an actual outcome, and never a superposition of potential outcomes. Thus, one cannot 'directly observe' potentiality, but rather only infer it from the structure of the theory.[4]

In what follows, we elaborate this basic metaphysical picture and discuss how it can help to make sense of quantum nonlocality, entanglement, and other related non-classical concepts that appear to be forced on us by quantum theory. It should be noted, however, that the authors have varied approaches to fleshing out the metaphysics in specific terms. Thus, the proposed metaphysical framework can be exemplified via alternative, but fundamentally compatible, formulations.

**II. Possibilist Realism vs. Actualism in Quantum Theory**

We first note that *res potentia* can be understood as a general concept applying to a broad range of possibilities. Traditionally, possibilist realism has encompassed all sorts of conceivable possibilities, as in some versions of modal logic, (e.g. Lewis 1986), and we do not advocate the broadest scope of the possibilities considered for realism.[5] We are primarily concerned with proposing that quantum entities and processes are a particularly robust subset of these, which we will call *quantum potentiae* (QP); and that these are strong candidates for realism. However, before focusing specifically on QP, let us first take note of an apparently mundane but ontologically significant aspect of the interplay between actualities and possibilities: namely, the way in which actual events can instantaneously and 'acausally' (in the sense of classical, efficient causality) alter what is next possible globally. As one of us (SK) has observed (Kauffman 2016, Chapter 7), we might plan to meet tomorrow for coffee at the Downtown Coffee Shop. But suppose that, unbeknownst to us, while we are making these plans, the coffee shop (actually) closes. Instantaneously and acausally, it is no longer possible for us (or for anyone no matter where they happen to live) to have coffee at the Downtown Coffee Shop tomorrow. What is possible has been globally and acausally altered by a new actual (token of res extensa).[6] In order for this to occur, no relativity-violating signal had to be sent; no physical law had to be violated. We simply allow that actual events can instantaneously and acausally affect *what is next possible* (given certain logical presuppositions, to be discussed presently) which, in turn, influences what can next become actual, and so on. In this way, there is an acausal 'gap' between res extensa and res potentia in their mutual interplay, that corresponds to a form of global nonlocality.[7] One might object that in the above example of ordinary macroscopic processes, the nonlocality seems confined to the influence of actuality on what is next possible, since in the apparently deterministic, classically conceived macroscopic world, actuals lead deterministically to new actuals (and 'what is possible' plays no real dynamical role). However, at the quantum level, this does not hold, so that the acausal gap really does exist in both directions (from actuals to possibles, and vice versa).

---

[3] cf. Epperson (2013, 86-87).

[4] We note that De Ronde (e.g. 2015) has also proposed ontological potentiae in connection with quantum theory.

[5] For example, we do not wish to assert categorically that 'It is possible that there are Aliens' given a world in which evolutionary processes have never yielded a race of beings called 'Aliens' (Mentzel 2016).

Moreover, we will see (in more detail below) that quantum potentiae (QP, represented by the usual quantum state or ray in Hilbert Space), like *res potentiae* in general, satisfy neither the Principle of Non-contradiction (PNC), nor the Law of the Excluded Middle (LEM), both formerly considered as 'self-evident' first principles of logic. Together, PNC and LEM constitute a principle of exclusive disjunction of contradictories, wherein a proposition P is necessarily either true or false, with no 'middle' alternative. Russell presented LEM this way: "Everything must either be or not be" (Russell, 1912, 113). When interpreted classically, Russell's formulation implicitly only acknowledges one mode of 'being'—that which is actual. Thus, a tacit classical assumption behind LEM is that of actualism: the doctrine that only actual things exist. However, as will be demonstrated presently, in the context of quantum mechanics, PNC and LEM together evince the ontological significance of *both* actuality and potentiality, given that every quantum measurement entails the former's

---

[6] While 'acausal' in the classical sense of efficient causality (wherein one actual state causally influences another actual state), in the quantum mechanical sense of causality wherein potentia are treated as ontologically significant, the actualized state is understood to 'causally' alter the probability distribution by which the next 'possible' state is defined. For further discussion of this distinction between classical efficient causality and quantum mechanical causality, see Epperson (2004, 92-93; 2013, 105-6). On the other hand, under certain circumstances and at the relativistic level, where decay probabilities are taken into account, the relation between an actualized state and the next QP state may itself be indeterministic (see, e.g. Kastner 2012, Section 3.4 and Chapter 6).

[7] cf. Epperson (2013, 60-62).

evolution from the latter by way of probabilities, which also satisfy both PNC and LEM.

For purposes of this discussion, we will assume that quantum measurements yield actual results in an indeterministic manner—i.e., one that cannot be reduced to the strictures of classical efficient causality. This indeterministic process is represented by the non-unitary von Neumann 'Process 1' measurement transition (Von Neumann, 1955, 352). Here, we differ from the usual assessment that the measurement problem remains unsolved, as expressed for example by R. Griffiths (2017). In fact, insofar as the measurement transition of von Neumann (from a pure to mixed state) is the primary aspect of the measurement problem, two of us (RK and ME) have proposed solutions, each unique in approach yet similarly grounded in the ontological interpretation of quantum potentiae proposed here.[8]

We further note that 'Process 1' can be broken down into two stages: (i) the transition from a pure state to a mixed state, which comprises N outcomes within a well-defined Boolean probability space; and (ii) the 'collapse' to one specific outcome (actual) with its associated probability (RTI accounts for (ii) in terms of a generalization of spontaneous symmetry breaking; i.e., one which includes the Born Rule weights). We will term the set of possible outcomes in step (i) as 'probable outcome states' to denote that they come with well-defined Kolmogorov probabilities. Probable outcome states of (i), like the actual outcome states resulting from step (ii), do satisfy PNC and LEM; therefore, it is the evolution of potential (pure) states to probable states that bridges *res potentia* and *res extensa* in quantum mechanics, such that the latter can be formalized as actualizations of the former. An example of a potential (pure) state is that of an electron bound within a hydrogen atom.

In what follows, we question the assumption of actualism and its consequence of LEM. First, it is a fundamental feature of quantum mechanics that the object of observation is always a macroscopic phenomenon; i.e., a detector click or the position of a pointer. That observation *indirectly,* but reliably,[9] allows an inference that the prepared quantum system now occupies an actual outcome state, as opposed to a superposition of pure states (the latter being forms of QP). Thus, one cannot 'directly observe' potentiality, but can infer it as a *calculably measurable* (not *observably measurable*) aspect of the quantum ontology. We believe that the latter has been overlooked in standard approaches to interpreting quantum theory, which presuppose actualism. In this regard, it may be useful to recall Ernan McMullin's important observation:

> …imaginability must not be made the test for ontology. The realist claim is that the scientist is discovering the structures of the world; it is not required   in

---

[8] Cf. Kastner (2012), (2016a), (2017a); Epperson and Zafiris (2013); Epperson (2004), (2009).
[9] We say 'reliably' because an outcome can be corroborated as veridically resulting in a specific quantum state.

> addition that these structures be imaginable in the categories of the macroworld. (McMullin 1984, 15)

The relevance of this remark in our present context is that, as creatures immersed in world of phenomena (which, on an individual level, are our sensory experiences of actual outcomes), it is easy for us to assume that the phenomena are the same as the ontology. Moreover, it is difficult for us to imagine or conceptualize any other categories of reality beyond the level of actual—i.e., what is immediately available to us in perceptual terms. Indeed, for millennia, focusing solely on the phenomena was absolutely necessary for survival; were we spending significant amounts of time imagining and conceiving of underlying and nonperceived realities, we would never have survived the process of evolution to our present state! Thus, our 'default setting' is actualism; but McMullin reminds us that this is not obligatory, and can serve as an impediment to progress in understanding.

In particular, one of us, RK, has criticized the prevailing actualist assumption in the context of certain 'time-symmetric' interpretations of quantum mechanics (such as those of Sutherland, Price and Wharton cited above) that everything exists within "the spacetime theater," which leads to a static block world ontology---yet one which is often portrayed as involving some sort of dynamical story (Kastner, 2017). Another of us, ME, with Elias Zafiris, has formalized a similar argument against the time-symmetric, actualist classical block world ontology, proposing a topological interpretation of quantum mechanics whereby spatiotemporal extensiveness and its metrical structure is emergent from dynamical topological quantum event structures and, respectively, the set theoretic framework of the former is generalized to the category theoretic framework of the latter (Epperson & Zafiris, 2013; see also Zafiris & Mallios, 2011).

Both of these arguments evince that a static block world comprising no more than a set of actual events cannot really be a dynamical ontology. This inconsistency is generally either simply ignored, or is glossed over by equivocation between ontological and epistemic considerations. A notable and creditable exception is the explicit acknowledgment by Stuckey, Silberstein *et al* that the 3+1 block world ontology is *adynamical* (e.g., Stuckey, Silberstein and Cifone 2008).

We can retain a truly dynamical account of quantum mechanics by taking into account *res potentia*. Feynman famously re-derived the quantum laws in his 'sum over paths' approach by taking a quantum system as taking 'all possible paths' from an initial prepared position to a final detected position (the latter constituting the result of a measurement). Clearly, such a system does not *actually* take distinct and mutually exclusive paths; its 'taking of all possible paths' is properly regarded as a set of possibilities, not actualities. Thus, Feynman's possible paths of a quantum entity exemplify our notion of *res potentia*; and his derivation of quantum theory implicitly rejects actualism. We suggest that the efficacy of his possibilist approach in yielding a formalism that was initially arrived at by heuristic mathematical data-fitting is evidence that it captures some ontological feature(s) of reality.

Thus, we propose that quantum mechanics evinces a reality that entails both actualities (*res extensa*) and potentia (*res potentia*), wherein the latter are as ontologically significant as the former, and not merely an epistemic abstraction as in classical mechanics. On this proposal, quantum mechanics IS about what exists in the world; but what exists comprises both possibles and actuals. Thus, while John Bell's insistence on "beables" as opposed to just "observables" constituted a laudable return to realism about quantum theory in the face of growing instrumentalism, he too fell into the default actualism assumption; i.e., he assumed that to 'be' meant 'to be actual,' so that his 'beables' were assumed to be actual but unknown hidden variables. Thus, the option of considering *potentiae* as something eligible for 'beable' status continued to be overlooked.

Regarding Feynman's reconstruction of quantum theory by way of 'possible paths': here it is worthwhile to recall Einstein's distinction between "constructive theories" and "principle theories." A constructive theory, according to Einstein, was one that built up the theory from basic physical concepts, in such a way that one could see physical processes at work in yielding the phenomena. That is, it provided a specific physical model. An example is the kinetic period of gases, which provided a constructive model of processes resulting in the empirical laws of thermodynamics. Einstein commented of constructive theories that:

> They attempt to build up a picture of the more complex phenomena out of the materials of a relatively simple formal scheme from which they start out. Thus the kinetic theory of gases seeks to reduce mechanical, thermal, and diffusional processes to movements of molecules – i.e., to build them up out of the hypothesis of molecular motion. When we say that we have succeeded in understanding a group of natural processes, we invariably mean that a constructive theory has been found which covers the processes in question. (Einstein, 1919)

In contrast, a principle theory was one that followed from one or more abstract principles, such as conservation laws, symmetries, etc. Quantum theory actually started out with Heisenberg as empirical data fitting, from which he obtained his matrix mechanics. Then Schrödinger brought into play specific principles, such as replacing energy and momentum with their space-time operators. It was actually not until Feynman that some form of constructive quantum theory was presented, however strange the proposed model was. Can one view the idea of an electron simultaneously pursuing all possible paths as a "model"? Certainly not in the classical, actualist sense we are used to. Nevertheless, it *is* a model, and it does yield the theory that was arrived at earlier through data fitting and abstract principles. We believe that Einstein's insight was correct – that when one has a constructive model, one gains insight into physical processes underlying the phenomena that one lacks with a principal-only theory. This leads us to consider the ontological reality of possibilities.

**III. Res potentia and Res extensa: Linked Through Measurement**

In this section, we discuss in more detail the key features of res potentia as embodied in the quantum potentiae (QP), the manner in which res potentia in the form of QP is transformed in res extensa through measurement, and implications for the relationship between QP and res extensa.

Consider the following proposition concerning a two-slit experiment:

*X. "The photon possibly went through slit A."*

Note that one can say of X: "X is true AND 'not X' is true" without contradiction. Thus X, as a statement of possibility, does not obey the law of the excluded middle. On the other hand, consider Y:

*Y. "The photon was detected at point P on the detection screen."*

Y, as a statement about an actuality, does obey the law of the excluded middle.

Proposition X applies to a situation involving a quantum superposition (an instance of Feynman's 'sum over paths'), while Y applies to the result of a measurement. Thus, we propose that measurement is a real, physical process, albeit indeterministic and acausal, that transforms possibles into actuals. In terms of our proposed non-substance dualism, *res potentia* is transformed into *res extensa* through measurement.

One specific, quantitative model of such a transformation is provided in the Relativistic Transactional Interpretation (RTI) (Kastner 2012, Chapters 3 and 6, and Appendix C on the specific conditions yielding measurement in the EPR context); another is given in the sheaf theoretic, topological Relational Realist (RR) interpretation (Epperson & Zafiris, 2013). One need not subscribe to either of these models in order to consider the current proposal, which simply points out in general terms the efficacy of allowing for a non-substance duality of *res potentia* and *res extensa*, where the former is transformed into the latter through measurement. Further, the concept of quantum mechanical actualization of potentia via measurement need not commit one to a specific theory of measurement itself (although we assume that measurement is genuinely non-unitary and that there are no hidden variables).

Consider again the two-slit experiment discussed above. If we wished, we could modify our experiment such that the measurement outcome triggered generation of a new quantum state (e.g., a photon prepared in a known pure state and subject to further measurement). In such a case, the measurement acausally yields a new actual (the outcome leading to the new prepared state), which in turn can bring about new quantum possibles (QP)—since the prepared pure state is a potentiality only. Since the bringing about of the new QP in this manner is not a causal process (it is indeterministic), *actuals (arising via measurement) acausally dictate what is next* *possible*. With this in mind, we may next see how allowing for this real interplay of res potentia with res extensa can help to make sense of some notorious peculiarities of quantum theory.

**III A. Possibles, Nonlocality, and Entanglement**

The quantum system pursuing "all possible paths" in Feynman's model is obviously engaging in a radically nonlocal activity: it is in all possible places at once at any given time. Of course, Feynman was working with a particle–like picture; in the wave picture, such nonlocal activity seems more natural, since a wave is naturally 'spread out'. However, the de Broglie waves corresponding to quantum states are not spacetime objects; it is only discrete, localized phenomena that are in-principle-observable elements of spacetime. In terms of our non-substance dualism, the de Broglie waves are the possibilities (res potentia), while the discrete localized phenomena are the actualities (res extensa). A possibility is, in principle, not a spacetime object; it is rather a *vehicle of enablement* (noncausal and inherently indeterministic*)* of spacetime actualities*.* Thus, a quantum entity, prior to actualization, is a nonlocal object (quantum potentia, QP). With this picture in mind, a composite system of more than one quantum entity, such as an entangled pair in an EPR experiment, can be seen as a naturally nonlocal QP form that can give rise to two actualities (i.e. two observable spacetime outcomes) instead of a single one.

As an example, consider the form of QP consisting of two entangled spin-1/2 degrees of freedom in a singlet state, with opposite momenta. Either can potentially be 'up' or 'down' along any measurement direction, but they are anticorrelated. If one is measured and found 'up,' that constitutes a new actual (i.e., a spacetime event and token of *res extensa*). Instantaneously and acausally, the possibility that the other will be measured to be 'up' (along the same axis) has vanished.[10] But nothing has disappeared from spacetime, nor has there been any influence exceeding the speed of light within spacetime; so there has been no violation of relativity, which governs only the domain of actuals (spacetime events). Notice also that, in keeping with relativity, there need be no fact of the matter about which degree of freedom is measured 'first.' The QP consisting of the two correlated degrees of freedom, not being a spacetime object, constrains--from beyond spacetime—the sets of events that can be actualized in spacetime.

What the EPR experiments reveal is that while there is, indeed, no measurable nonlocal, efficient causal influence between *A* and *B*, there *is* a measurable, nonlocal probability conditionalization between *A* and *B* that always takes the form of an asymmetrical internal relation. For example, given the outcome at *A*, the outcome at *B* is internally related to that outcome. This is manifest as a probability conditionalization of the *potential* outcomes at *B* by the *actual* outcome at *A*. When considering the phrase "given the outcome at A," however, it is crucial to distinguish between 'logical antecedence' and 'temporal antecedence' here, for these are often unreflectively assimilated. Temporal antecedence refers to an asymmetrical *metrical coordination of objects* according to the parameter of time (or more accurately, spacetime). In contrast, logical antecedence refers to an asymmetrical *logical supersession of events* such as that implied by the notion of conditional probability or more broadly, propositional logic; but in a quantum mechanical context. In particular, the novelty of the quantum formalism is that (as applied to the EPR case) it consistently integrates the asymmetrical dependencies P(B|A) and P(A|B), reflecting that there need be no fact of the matter about which outcome is taken as 'given first.' Specifically, one takes into account the observables being measured locally at A and B—which transform possible outcomes to probable outcomes as in the above terminology—and those together (regardless of temporal order) dictate the Boolean probability space applying to the probable outcomes.[11]

---

[10] It is often supposed that measurement results obtain as a result of decoherence in a unitary-only account, which leaves small but nonvanishing off-diagonal elements; but that approach does not solve the problem of measurement and is arguably circular (cf. Kastner 2014). Again, here we assume that measurement is a non-unitary process corresponding to von Neumann's 'Process 1,' which really does yield an exactly diagonal density matrix that can be interpreted epistemically. Thus, upon actualization of one outcome (step (ii) of the measurement transition, 'collapse'), the others have truly vanished.

We thus propose that allowing for the dualism of res potentia/res extensa can serve to explain non-local phenomena. It can do so by observing that the phenomena are indeed correlated (through their supporting potentiae), but not causally connected in the usual way. That is, there is no efficient causal interaction between actuals; so we need not be concerned with the limitation of the speed of light on 'signals' between the two wings of the EPR pair (of which there are none), nor do we need to invent hidden variables that are not in the theory itself, or invoke never-observed exotic particles such as tachyons (Maudlin 2011. p. 71).

**III B. Instantaneous, nonlocal change in the wavefunction upon measurement**

The same basic point holds for the more general case of N entangled spins, N ≥ 2. If one is measured and found 'up,' instantaneously the wave function for the remaining N -1 degrees of freedom changes. It is a mystery in standard actualist approaches to quantum mechanics how this can be the case. But in the ontology of res potentia/res extensa, the explanation is straightforward: the result of the measurement of the first particle is a new actual that instantaneously and acausally alters what is next possible, just as in the example above with the Downtown Coffee Shop. In this case, 'what is possible' *is* just the state of the remaining N-1 degrees of freedom.

---

[11] For additional elaboration, see Epperson (2013, 71-80) and Kastner (2012, Appendix C). The latter gives a physical account of how local measurements dictate the relevant probability space. Though in the specific experiment there happens to a temporal order due to timelike separation of the measurements, the same process applies regardless of temporal order.

As another example, consider photons in a two-slit experiment. If they are measured to have gone through either the left or right slit, the interference pattern disappears. Why? Prior to measurement, both possibilities corresponding to passage through the left and right slit exist. If a new actual occurs, via measurement, regarding passage through the left (or right) slit, the "possibility of passage through the right (or left) slit" vanishes, and with it the interference of the two quantum potentiae (QP), one of which no longer exists.

The same basic process explains the phenomenon of 'null measurement': if, in the two-slit experiment, the photons are measured to have not gone through the left slit (i.e., NOT-L becomes a new actual), then (since actuals obey the law of the excluded middle), they can only have gone through the right slit. The interference pattern, which can only arise if QP for passage through both slits are really present, therefore vanishes.

Thus, we propose that an ontological dualism of res potentia/res extensa affords an account of quantum non-locality, instantaneous and global wave function changes for N entangled spins when one is measured, "which-way information" corresponding to loss of interference, and the phenomena associated with null measurements. These are all key puzzling aspects of standard quantum theory that are not readily explained otherwise. Admittedly, this requires expanding our ontology beyond the merely 'actual'; but we believe that it is time to do so, given that many researchers are tacitly, or even explicitly, making use of Heisenberg's idea that quantum systems are forms of potentiae, and/or that what goes on in spacetime may not be the entire ontological story. We return to this point in section V.

**IV. Potentiae Beyond Quantum Mechanics?**

We have been focusing above on quantum potentiae (QP). But, returning to our "Downtown Coffee Shop" example in Section 1, "possibilities" are broadly used in normal life and are the subject of modal logic, which is not specifically geared to the quantum level. How broadly should we take potentiae to be "real"?

It's useful here to make a quantitative distinction between *quantum potentiae* (QP) and other, even less 'substantial' forms of *res potentia* (call the latter RP). First, the QP are in-principle *quantifiable*: given a quantum state such as |Z+> ("spin up along z"), one can define in precise quantitative terms all the actualities that may result, depending on the choice of measurement. In addition, given an actual measurement process, one of the probable outcomes defined by the QP will indeed result: all measurements will have an outcome. This is quantified by the three probability axioms of Kolmogorov. Most significant in this context is the axiom of unit measure (AUM): the probability that at least one of the outcomes corresponding to an eigenvalue of the observable being measured on the given QP will be actualized is 1.

In contrast, an example of the more general RP is the non-quantifiable set of possibilities that may be enabled by a specific evolutionary development, such as the advent of the swim bladder in certain species of fish. One of us, SK, calls the set of resulting new possibilities "unprestatable," signifying that there is no way, even in principle, to define or enumerate such a set of possibilities (Kauffman 2012, Chapter 4). Once such an organ becomes actual, an indefinite number of unprestatable new possibilities is enabled. For example, a parasitic worm could take up residence in the swim bladder; a disease of the swim bladder could result in fish swimming upside down (which is already an actuality); a swim bladder could develop into some other organ. As for these general sorts of RP, they could all be actualized; or some could be, or none at all. Thus, such possibilities do not obey the probability axioms of Kolmogorov, and in that sense are not quantifiable.

Yet clearly, such possibilities are enabled when a new actuality occurs, and vice versa (new actualities may arise from the new possibilities). It seems true that once the swim bladder evolves, it is really true that a worm might evolve to live in swim bladders. Thus, we confront in evolution and aspects of normal life what appear to be real possibilities that are not quantifiable. They are indefinite. Unlike the usual cases involving probability, the sample space is not known or even defined. Not only do we not know what will happen, we do not even know what *can* happen. Whether and how the quantum *res potentia* we here advocate may relate to what seem to be the real but open-ended potentia of biological evolution is as yet unclear, but worthy of further inquiry.

**V. Are Potentiae Outside Spacetime?**

Returning to the more substantial, quantifiable QP (such as a quantum system described by state |Z+> ): these are pre-actual modes of being, and as such, they are not elements of spacetime. Such an entity is a necessary but not sufficient condition for an actuality that is housed in spacetime; the sufficient condition is that measurement of the entity occur.[12] In this perspective, nonlocal correlations such as those of the EPR experiment can be understood as a natural, mutually constrained relationship between the kinds of spacetime actualities that can result from a given possibility—which itself is not a spacetime entity.

This new ontological picture requires that we expand our concept of 'what is real' to include an extraspatiotemporal domain of quantum possibility. Thus, we need to 'think outside the spacetime box.'. Other researchers have recently suggested that spacetime is not fundamental. For example, Ney has been advocating what she terms "Wave Function Realism," in which the wave function is taken as ontologically real and spacetime phenomena comprise only a subspace of that ontology: "What appears in the derivative three-dimensional metaphysics as nonlocal influence is explained by the evolution of the wave function in its space where there are no nonlocal influences." (Ney 2017). Our approach differs in that we regard measurement as a real, non-unitary process, and do not take the universe as a whole to be described by a position-basis wavefunction; but the spirit of allowing for a larger ontology for the quantum realm is essentially the same.

---

[12] According to both RTI and RR, the existence of such a state is contingent on there being a measurement context, defined by the relevant Hamiltonian and (a) absorber response(s) in the case of RTI, or (b) sheaf of Boolean reference frames in the case of RR. Thus, such QP are clearly more robust than the general RP discussed above: once they exist, something actual *will* occur.

Another important example is a set of remarks by Anton Zeilinger considering the enigma of entanglement. Zeilinger notes that:

> ..it appears that on the level of measurements of properties of members of an entangled ensemble, quantum physics is oblivious to space and time.
>
> It appears that an understanding is possible via the notion of information. Information seen as the possibility of obtaining knowledge. Then quantum entanglement describes a situation where information exists about possible correlations between possible future results of possible future measurements without any information existing for the individual measurements. The latter explains quantum randomness, the first quantum entanglement. And both have significant consequences for our customary notions of causality.
>
> It remains to be seen what the consequences are for our notions of space and time, or space-time for that matter. Space-time itself cannot be above or beyond such considerations. I suggest we need a new deep analysis of space-time, a conceptual analysis maybe analogous to the one done by the Viennese physicist-philosopher Ernst Mach who kicked Newton's absolute space and absolute time form their throne. (Zeilinger 2016)

What the current authors are endeavoring to do, in fact, is just that: to 'kick off the throne' the usual conception of spacetime as an all-encompassing container for all that exists. What Zeilinger refers to above as 'information' we suggest should be understood as ontologically real, pre-spacetime possibilities—since clearly they are doing something that constrains the actualities of our experience. We would caution against taking the idea of 'information' as an observer-dependent, epistemic notion, which forecloses a realist understanding of the quantum formalism, and compromises our ability to subject spacetime to the critical considerations that Zeilinger rightly suggests it should be.

## VI. Conclusion

We have argued that an appropriate realist understanding of quantum mechanics calls for the metaphysical category of *res potentia*, just as Heisenberg suggested long ago. In particular, we suggest a non-substance dualism of res potentia and res extensa as mutually implicative modes of existence, where quantum states instantiate a particular, quantifiable form of res potentia, 'Quantum Potentiae' (QP). As non-actuals, QP are not spacetime objects, and they do not obey the Law of the Excluded Middle (LEM) or the Principle of Non-Contradiction (PNC). On the other hand, *res extensa* is exemplified by the outcomes of measurements, which constitute structured elements of spacetime; the latter, as actuals, obey LEM and PNC. We argue that measurement is a real physical process that transforms quantum potentiae into elements of res extensa, in a non-unitary and classically acausal process, and we offer specific models of such a measurement process. In this ontology, spacetime (the structured set of actuals) emerges from a quantum substratum, as actuals 'crystallizing' out of a more fluid domain of possibles;[13] thus, spacetime is not all that exists.

The above picture accounts naturally for the counter-intuitive features of quantum mechanics such as nonlocality, entanglement, and instantaneous collapse. We affirm Zeilinger's call for critical examination of the usual notion of spacetime as a fundamental domain for all that exists, and urge that this is what needs to be dropped in order to make progress in understanding what our best physical theories may be telling us about Nature.

Acknowledgments. The authors are grateful to Robert B. Griffiths and Christian de Ronde for valuable correspondence, and to an anonymous referee for suggestions for improvement of the presentation.

---

[13] Cf. Kastner (2016b) for the connection of spacetime emergence with causal sets.